# Terahertz sensing of 7nm dielectric film with bound states in the continuum metasurfaces

**Yogesh Kumar Srivastava,[1,2] Rajour Tanyi Ako,[3] Manoj Gupta,[1,2] Madhu Bhaskaran,[3] Sharath Sriram[3] and Ranjan Singh[1,2,*]**

*[1]Division of Physics and Applied Physics, School of Physical and Mathematical Sciences, Nanyang Technological University, 21 Nanyang Link, Singapore 637371*

*[2]Centre for Disruptive Photonic Technologies, The Photonics Institute, Nanyang Technological University, 50 Nanyang Avenue, Singapore 639798*

*[3]Functional Materials and Microsystems Research Group and the Micro Nano Research Facility, RMIT University, Melbourne, VIC 3000, Australia*

*Email: ranjans@ntu.edu.sg

## Abstract

Fingerprint spectral response of several materials with terahertz electromagnetic radiation indicates that terahertz technology is an effective tool for sensing applications. However, sensing few nanometer thin-film of dielectrics with much longer terahertz waves (1 THz = 0.3 mm) is challenging. Here, we demonstrate a quasi-bound state in the continuum (BIC) resonance for sensing of nanometer scale thin analyte deposited on a flexible metasurface. The large sensitivity originates from strong local field confinement of the quasi-BIC Fano resonance state and extremely low absorption loss of a low-index cyclic olefin copolymer substrate. A minimum thickness of 7 nm thin-film of germanium is sensed on the metasurface, which corresponds to a deep subwavelength length scale of $\lambda/43000$, where $\lambda$ is the resonance wavelength. The low-loss, flexible and large mechanical strength of the quasi-BIC micro structured metamaterial sensor could be an ideal platform for developing ultrasensitive wearable terahertz sensors.

Potential applications of terahertz technology in a large number of industries such as biomedical, security screening and wireless communication have been demonstrated.[1-4] The most important applications of the terahertz technology is in non-ionizing, non-destructive, fast imaging and sensing modalities.[3] Various materials such as DNA, proteins, fungus, and explosives show rich spectral fingerprints dominated by their intramolecular and intermolecular vibrational modes in the terahertz regime of the electromagnetic spectrum, which indicates that the terahertz technology is a highly effective tool for the sensing applications.[5-9] Among different sensing techniques, the sensors based on metamaterials, which are periodic arrays of artificially designed resonant elements, show remarkable sensitivity to the small volume of analyte. This sensitivity arises due to enhanced light-matter interaction owing to strong confinement of electromagnetic fields in the sub-wavelength structure.[10-21] A sharp resonance feature is typically desired to detect minute frequency shift that arises due to low volume of the analyte.[22, 23] Moreover, the amplitude of the resonance should be strong enough so that it could be easily detected in the noisy environment. In this context, a variety of resonant structures such as metamaterials and dielectric resonator platforms possessing sharp spectral features like Fano, quadrupole, and toroidal resonances have been demonstrated for refractive index sensing.[9, 22-26] However, sensing nanometer thin dielectric layers with sub-millimeter scale terahertz waves is challenging since it typically requires nano-confined terahertz fields which involves highly complex fabrication scheme.[27] The fringing fields of the metamaterials with micron size split gaps at terahertz wavelength ($\lambda$) extend to about $\lambda/20$ (few microns) and play a vital role in sensing any changes in the dielectric environment of the metamaterials.[24, 28, 29] Thus, to harness complete sensing capability of metamaterial sensors at terahertz frequencies, usually large thickness of the analyte is desired.[25, 30, 31] On reducing the analyte thickness to few nanometers, the shift in the resonance frequency becomes less significant due to reduced strength of light-matter interaction. In the conventional terahertz time-domain spectroscopy (THz-TDS) measurements, such small shift in resonance frequency could only be measured with extremely sharp resonance features. This requires high spectral resolution which could only be obtained by longer temporal scans. Therefore, the frequency shift based sensing method becomes less effective for extremely low analyte volumes, and there arises a need for development of an alternative technique for sensing nanometer scale thin-film analytes at terahertz frequencies.

Here, we exploit quasi-bound state in the continuum (BIC) Fano resonance observed in metamaterial structures for sensing ultra-thin analyte layers.[32-35] An ideal BIC is strictly a mathematical concept which results in an infinite quality (Q) factor, and it only exists in an ideal metamaterial with highly confined non-radiative mode that does not couple to the free space.[32, 34, 35] Since an ideal BIC is not observable in real systems, we access a *quasi-BIC* mode through their exponentially diverging Q factor trajectory. Ideal BIC states are symmetry protected and can radiate to far-field in the form of quasi-BIC (super-cavity) mode by creating a leakage channel through symmetry breaking.[34, 35] It has been recently established theoretically that the concept of Fano resonance in symmetry broken split ring resonators is a special case of BIC, known as quasi-BIC.[34] In this work, we demonstrate a quasi-BIC (Fano) based terahertz metamaterial sensor fabricated on a flexible and, low refractive index (refractive index $n = 1.53$ and loss tangent $tan\delta \sim$ 0.004-0.0006 at 0.2-2.5 THz) substrate of cyclic olefin copolymer (COC) for sensing deep subwavelength thicknesses of analyte.[31, 36-39] This sensor renders large sensitivity owing to the low refractive index of the substrates combined with the strongly confined local fields of the quasi-BIC state in micron dimension capacitive gaps. Moreover, we utilize a recently proposed sensing analysis by Al-Naib based on the change in amplitude and phase of the broad transmission spectra due to the presence of sub-micron analyte overlayer on the top surface of the metamaterial.[21] The amplitude and phase difference based sensing is extremely fast and does not require any extra preparation steps. Using this method, we demonstrate sensing of extremely deep subwavelength analyte thickness of 7 nm which is ~1/43000 times of the metamaterial resonant wavelength. Furthermore, the COC substrates are robust and extremely flexible which makes them easily adaptable for the real world wearable photonic sensors.

In this study, we choose a typical double gap terahertz asymmetric split ring (TASR) as a building block for the metamaterial due to its design flexibility and sharp quasi-BIC Fano resonant spectral feature.[22, 35, 40-42] A small portion of a large array of two-dimensional planar TASR metamaterial fabricated on a COC substrate is shown in Figure 1(a). The geometrical dimensions of the unit cell are revealed in the inset of Figure 1(a). The symmetry of the structure is broken by moving one of the capacitive split gaps away from the central vertical axis which leads to the excitation of sharp quasi-BIC Fano resonance state due to distortion of symmetry protected BIC.[34, 35] The TASR arms act as an inductor while the electric field is confined and enhanced locally in the vicinity of the capacitive split gaps. This TASR metamaterial resides on a low refractive index COC substrate, which is a very promising dielectric material with excellent properties such as high chemical resistance, low water absorption, high flexibility, good thin film compatibility (with use of appropriate adhesion layers) and large transparency across terahertz (0.2-2.5 THz) region of the electromagnetic spectrum.[37, 43] The TASR structures are fabricated using standard photolithography technique followed by thermal evaporation of 200 nm thick gold (Au) layer with a 20 nm chromium (Cr) adhesion layer. Details of the fabrication method are given in section 2 of the supplementary material. The sensing performance is studied by thermally depositing different thicknesses of germanium (Ge) on the fabricated TASR metamaterials, which results in relatively large resonance frequency shift due to its high dielectric constant ($\varepsilon = 16$). Figure 1(b) reveals high flexibility and free-standing nature of fabricated metamaterials, so that it can be bent, rolled and twisted as per the application requirements. The properties of the quasi-BIC Fano resonance, i.e., $Q$ factor and amplitude of the resonance are determined by the structural asymmetry '$d$' values.[35, 40, 44] Figure 1(c) depicts the $Q$ factors of quasi-BIC Fano extracted from the simulated transmission spectra of the ideal metamaterial composed of perfect electrical conductor (PEC) and realistic metallic metamaterial array on COC substrate with varying asymmetry parameter $d$ (see supplementary material for details on numerical simulations and $Q$ factor extraction). The $Q$ factor diverges ($\infty$) at $d = 0$ due to no coupling of the mode to the free space. Breaking the structural symmetry by moving lower TASR gap away from the central vertical axis leads to the transition of BIC to the quasi-BIC state and the $Q$ factors decrease gradually with increasing asymmetry. The $Q$ factors of an ideal metamaterial are significantly larger than the realistic metamaterial owing to their zero ohmic loss. We chose an asymmetry of $d = 10$ μm as a quasi-BIC configuration of the Fano resonance for experimental demonstration to easily capture spectral footprints in the measured transmission spectra. The simulated transmission spectra of chosen TASR metamaterial at $d = 10$ μm is shown in the inset of Figure 1(c). To demonstrate sensing of nanometer scale thin-film of analyte using TASR metamaterial, we first performed numerical simulations by placing an analyte of dielectric constant, $\varepsilon = 16$ on the metamaterial and gradually varied its thickness from 7 nm to 40 nm. Sensing performance of the metamaterial is evaluated using the proposed differential technique where the transmission amplitude and phase of the TASR metamaterial with analyte were subtracted from the transmission amplitude and phase without analyte, respectively.[21] Figure 1(d) and 1(e), respectively, show the change in the transmission amplitude ($\Delta T$) and phase ($\Delta \phi$) with different thicknesses of Ge overlayer ranging from 7 nm to 20 nm. We observed a significant change in the transmission amplitude and phase even for 7 nm thick analyte (shown by red curve), which illustrates the effectiveness of this method for the ultrasensitive sensing of deep subwavelength thicknesses of the analyte.

In order to explicitly present the importance of proposed sensing technique over conventional sensing methods, we carried out detailed simulations as well as experiments on the TASR metamaterials with different thicknesses of Ge. Terahertz transmission spectra of the fabricated metamaterials are obtained using fiber-based THz-TDS system. The terahertz beam spot of diameter, 7 mm, was illuminated on the metamaterial array of size, 20 mm × 20 mm. We performed 320 ps time scan to achieve a frequency resolution of about 3 GHz. Terahertz transmission amplitude is obtained by using the equation, $\tilde{T}(\omega) = \frac{\tilde{E}_S(\omega)}{\tilde{E}_R(\omega)}$, where $\tilde{E}_S(\omega)$ and $\tilde{E}_R(\omega)$ are the Fourier transformed electric field signal through the sample and reference, respectively. Figure 2(a) and (b), depicts the simulated and experimentally measured transmission spectra of the TASR metamaterial without (black, dotted curve) and with 7 nm thick Ge (red, solid curve) overlayer, which shows a red shift of 6 GHz and 3 GHz, respectively, for the quasi-BIC Fano resonance. On increasing the Ge thickness to 20 nm and 40 nm, the respective simulated spectrum shows a red shift of 14 GHz and 29 GHz, as shown in Figure 2(c) and 2(e). Figure 2(d) and 2(f) depicts experimentally measured transmission spectra of TASR metamaterial with 20 and 40 nm thick Ge overlayer which shows a red shift of 6 GHz and 22 GHz, respectively, as compared to the uncoated metamaterial. The increased thickness of analyte results in the enhanced light-matter interaction thus leading to the gradual red shifting of quasi-BIC resonance frequency. Here it is noted that the red shift in quasi-BIC Fano state is very small at low analyte thickness. Such a small shift in resonance frequency could be accurately measured only by increased time scan for the terahertz pulse in the THz-TDS measurements. Therefore, the conventional sensing measurement method becomes inefficient for sensing nanometer scale thin analyte at terahertz frequencies.

To overcome the difficulty in sensing extremely thin analyte films using the conventional THz-TDS method, we utilized an amplitude and phase difference method in which the terahertz transmission amplitude and phase of the TASR metamaterial with analyte were subtracted from the transmission amplitude and phase of TASR metamaterials without analyte.[21] Figure 3(a) represents the change in simulated transmission amplitude ($\Delta T$) of TASR metamaterial in the presence of analyte overlayers of different thicknesses with respect to the TASR response without analyte. On increasing the thickness of the analyte, $\Delta T$ increases gradually. The experimental values of $\Delta T$ for various thicknesses of the Ge overlayer are presented in Figure 3(b). The simulated curves show peak-to-peak amplitude difference ($|\Delta T|$) value of 0.11 for 7 nm analyte while corresponding experimental results show the $|\Delta T|$ value of 0.05. The observed $|\Delta T|$ values are quite significant and could be easily detected with terahertz transmission measurements in a noisy environment. The small difference between the simulated and experimentally measured values are due to the limited resolution of the THz-TDS system and the quality of the fabricated metamaterial sample. Similarly, Figure 3(c) and 3(d), respectively, depict the simulated and experimentally measured change in phase ($\Delta\phi$) of TASR metamaterial without and with the analyte of thicknesses ranging from 7 nm to 40 nm. We observed that the measured results (Figure 3(d)) are in agreement with the corresponding simulations performed using CST Microwave studio. The peak-to-peak values of $\Delta\phi$ are 36.5º and 9.2º for the 7 nm thick analyte in simulation (Figure 3(c)) and measurement (Figure 3 (d)), respectively. The extracted $\Delta\phi$ values are quite significant and could be easily detected in the THz-TDS measurements. Therefore, monitoring the amplitude and phase difference enables sensing of extremely thin-film analytes.

We further analyzed sensitivity of the TASR metamaterials residing on COC substrate by placing an analyte of constant thickness of 40 nm but varying the refractive index and compared their sensitivity with the identical TASR metamaterials residing on standard flexible Kapton (dielectric constant  $\varepsilon$ = 3.7936) substrate which is one of the most widely used flexible substrates. We defined the sensitivity of this refractive index sensor as the peak-to-peak transmission amplitude difference on coating analyte on metamaterial, per unit change in the refractive index of the analyte given by $= \frac{\Delta(|\Delta T|)}{\Delta n}$, where $|\Delta T|$ is the peak-to-peak transmission amplitude difference due to analyte overlayer and $\Delta n$ is the change in refractive index. Figure 4(a) and (b) depicts the transmission amplitude difference on varying refractive index of the 40 nm thick analyte overlayer placed on the TASR metamaterial residing on COC and Kapton substrates, respectively. Peak-to-peak amplitude difference of 0.83 and 0.49 is observed on varying the refractive index from $n = 1$ to $n = 4$ for COC and Kapton substrates, respectively. We plotted peak-to-peak transmission amplitude difference ($|\Delta T|$) with the change in refractive index of the analyte placed on TASR metamaterials residing on COC and Kapton substrates in Figure 4(c) and estimated the sensitivity of metamaterials for both substrates. The estimated sensitivity of TASR metamaterial residing on COC substrate as a function of the refractive index unit (RIU) turns out to be 0.28/RIU and it reduces to 0.16/RIU for the metamaterial residing on Kapton substrate. Therefore, the proposed TASR metamaterial on COC substrates show about 1.75 times higher sensitivity to the dielectric environment in comparison to the TASR metamaterials residing on Kapton substrates. Similarly, $\frac{\Delta(|\Delta \phi|)}{\Delta n}$ is about 5.68 times larger for COC substrate in comparison to Kapton substrate (section 5 of supplementary material). Further, we deposited 40 nm thick Ge on the fabricated TASR metamaterial residing on COC and Kapton substrates and estimated the transmission amplitude difference with respect to corresponding uncoated metamaterials as shown in the inset of Figure 4(a) and 4(b), respectively. We observed $|\Delta T|$ of 0.27 and 0.10 for metamaterials on COC and Kapton substrates, respectively, which further verify higher sensitivity of TASR metamaterial on COC substrates and thus could be an ideal platform for sensing deep subwavelength thicknesses of the analyte.

In summary, we have demonstrated an ultrasensitive terahertz quasi-BIC Fano resonant metamaterial sensor fabricated on an extremely low-loss flexible COC substrate which enables sensing of 7 nm thick analyte at terahertz frequencies. Moreover, our results validate high-performance sensing approach based on the change in amplitude and phase of metamaterial response without and with analyte overlayer to effectively sense deep subwavelength thicknesses of the analyte thin-films. The proposed sensor could find potential applications in wearable technologies for human health monitoring and extremely low volume detection of biological and chemical analytes.

**Supplementary Material**

See supplementary material for the information on numerical simulations, sample fabrication, analyte deposition, quality factor extraction of the quasi-BIC Fano resonances and electric field distribution at the metamaterial split gap.

**Acknowledgments**

The authors acknowledge valuable and timely assistance from Zhang Qiannan in performing thickness measurements of analyte layer using Atomic Force Microscopy. YKS, MG and RS acknowledge the research funding support from the Ministry of Education AcRF Tier 1 grant RG191/17 and Tier 2 grant MOE2017-T2-1-110. SS and RS acknowledge support from a RMIT Foundation Research Exchange Fellowship. This work was performed in part at the Micro Nano Research Facility at RMIT University in the Victorian Node of the Australian National Fabrication Facility (ANFF).

**References:**

1.      M. Tonouchi, Nature Photonics **1**, 97 (2007).
2.      B. Ferguson and X.-C. Zhang, Nature Materials **1**, 26 (2002).
3.      W. Withayachumnankul, G. M. Png, X. Yin, S. Atakaramians, I. Jones, H. Lin, B. S. Y. Ung, J. Balakrishnan, B. W. Ng, B. Ferguson, S. P. Mickan, B. M. Fischer and D. Abbott, Proceedings of the IEEE **95** (8), 1528-1558 (2007).
4.      M. C. Kemp, P. F. Taday, B. E. Cole, J. A. Cluff, A. J. Fitzgerald and W. R. Tribe, presented at the AeroSense 2003, 2003 (unpublished).
5.      J. Chen, Y. Chen, H. Zhao, G. J. Bastiaans and X. C. Zhang, Optics Express **15** (19), 12060-12067 (2007).
6.      H.-B. Liu, Y. Chen, G. J. Bastiaans and X. C. Zhang, Optics Express **14** (1), 415-423 (2006).
7.      B. M. Fischer, M. Walther and P. U. Jepsen, Physics in Medicine & Biology **47** (21), 3807 (2002).
8.      S. J. Park, J. T. Hong, S. J. Choi, H. S. Kim, W. K. Park, S. T. Han, J. Y. Park, S. Lee, D. S. Kim and Y. H. Ahn, Scientific Reports **4**, 4988 (2014).
9.      W. Xu, L. Xie and Y. Ying, Nanoscale **9** (37), 13864-13878 (2017).
10.     D. R. Smith, J. B. Pendry and M. C. K. Wiltshire, Science **305** (5685), 788 (2004).
11.     Y. Chen, I. A. I. Al-Naib, J. Gu, M. Wang, T. Ozaki, R. Morandotti and W. Zhang, AIP Advances **2** (2), 022109 (2012).
12.     N. Liu, M. Mesch, T. Weiss, M. Hentschel and H. Giessen, Nano Letters **10** (7), 2342-2348 (2010).
13.     K. V. Sreekanth, S. Sreejith, S. Han, A. Mishra, X. Chen, H. Sun, C. T. Lim and R. Singh, Nature Communications **9** (1), 369 (2018).
14.     H. Im, H. Shao, Y. I. Park, V. M. Peterson, C. M. Castro, R. Weissleder and H. Lee, Nature Biotechnology **32**, 490 (2014).
15.     C. Wu, A. B. Khanikaev, R. Adato, N. Arju, A. A. Yanik, H. Altug and G. Shvets, Nature Materials **11**, 69 (2011).
16.     N. Liu, T. Weiss, M. Mesch, L. Langguth, U. Eigenthaler, M. Hirscher, C. Sönnichsen and H. Giessen, Nano Letters **10** (4), 1103-1107 (2010).
17.     A. V. Kabashin, P. Evans, S. Pastkovsky, W. Hendren, G. A. Wurtz, R. Atkinson, R. Pollard, V. A. Podolskiy and A. V. Zayats, Nature Materials **8**, 867 (2009).
18.     N. Born, I. Al-Naib, C. Jansen, R. Singh, J. V. Moloney, M. Scheller and M. Koch, Advanced Optical Materials **3** (5), 642-645 (2015).
19.     J. F. O'Hara, R. Singh, I. Brener, E. Smirnova, J. Han, A. J. Taylor and W. Zhang, Optics Express **16** (3), 1786-1795 (2008).
20.     J. N. Anker, W. P. Hall, O. Lyandres, N. C. Shah, J. Zhao and R. P. Van Duyne, Nature Materials **7**, 442 (2008).
21.     I. Al-Naib, Journal of King Saud University - Science, https://doi.org/10.1016/j.jksus.2018.11.011.
22.     R. Singh, W. Cao, I. Al-Naib, L. Cong, W. Withayachumnankul and W. Zhang, Applied Physics Letters **105** (17), 171101 (2014).
23.     C. Wu, A. B. Khanikaev, R. Adato, N. Arju, A. A. Yanik, H. Altug and G. Shvets, Nat Mater **11** (1), 69-75 (2012).
24.     Y. K. Srivastava, L. Cong and R. Singh, Applied Physics Letters **111** (20), 201101 (2017).
25.     M. Gupta, Y. K. Srivastava, M. Manjappa and R. Singh, Applied Physics Letters **110** (12), 121108 (2017).
26.     I. Al-Naib,  IEEE Journal of Selected Topics in Quantum Electronics **23** (4), 1 (2017).
27.     H.-R. Park, X. Chen, N.-C. Nguyen, J. Peraire and S.-H. Oh, ACS Photonics **2** (3), 417-424 (2015).

28.     K. Shih, P. Pitchappa, L. Jin, C.-H. Chen, R. Singh and C. Lee, Applied Physics Letters **113** (7), 071105 (2018).

29.     W. Withayachumnankul, J. F. O'Hara, W. Cao, I. Al-Naib and W. Zhang, Optics Express **22** (1), 972-986 (2014).

30.     T. Suzuki, T. Kondo, Y. Ogawa, S. Kamba and N. Kondo, IEEE Sensors Journal **13** (12), 4972-4976 (2013).

31.     B. Reinhard, K. M. Schmitt, V. Wollrab, J. Neu, R. Beigang and M. Rahm, Applied Physics Letters **100** (22), 221101 (2012).

32.     C. W. Hsu, B. Zhen, A. D. Stone, J. D. Joannopoulos and M. Soljačić, Nature Reviews Materials **1**, 16048 (2016).

33.     C. W. Hsu, B. Zhen, J. Lee, S.-L. Chua, S. G. Johnson, J. D. Joannopoulos and M. Soljačić, Nature **499**, 188 (2013).

34.     K. Koshelev, S. Lepeshov, M. Liu, A. Bogdanov and Y. Kivshar, Physical Review Letters **121** (19), 193903 (2018).

35.     L. Cong and R. Singh, Advanced Optical Materials **0** (0), 1900383 (2019).

36.     L. M. Diaz-Albarran, E. Lugo-Hernandez, E. Ramirez-Garcia, M. A. Enciso-Aguilar, D. Valdez-Perez, P. Cereceda-Company, D. Granados and J. L. Costa-Krämer, Microelectronic Engineering **191**, 84-90 (2018).

37.     F. Pavanello, F. Garet, M.-B. Kuppam, E. Peytavit, M. Vanwolleghem, F. Vaurette, J.-L. Coutaz and J.-F. Lampin, Applied Physics Letters **102** (11), 111114 (2013).

38.     K. Nielsen, H. K. Rasmussen, A. J. L. Adam, P. C. M. Planken, O. Bang and P. U. Jepsen, Optics Express **17** (10), 8592-8601 (2009).

39.     F. D'Angelo, Z. Mics, M. Bonn and D. Turchinovich, Optics Express **22** (10), 12475-12485 (2014).

40.     L. Cong, M. Manjappa, N. Xu, I. Al-Naib, W. Zhang and R. Singh, Advanced Optical Materials **3** (11), 1537-1543 (2015).

41.     A. E. Miroshnichenko, S. Flach and Y. S. Kivshar, Reviews of Modern Physics **82** (3), 2257-2298 (2010).

42.     B. Luk'yanchuk, N. I. Zheludev, S. A. Maier, N. J. Halas, P. Nordlander, H. Giessen and C. T. Chong, Nature Materials **9**, 707 (2010).

43.     P. D. Cunningham, N. N. Valdes, F. A. Vallejo, L. M. Hayden, B. Polishak, X.-H. Zhou, J. Luo, A. K. Y. Jen, J. C. Williams and R. J. Twieg, Journal of Applied Physics **109** (4), 043505-043505-043505 (2011).

44.     Y. K. Srivastava, M. Manjappa, L. Cong, W. Cao, I. Al-Naib, W. Zhang and R. Singh, Advanced Optical Materials **4** (3), 457-463 (2016).

**Figure 1**

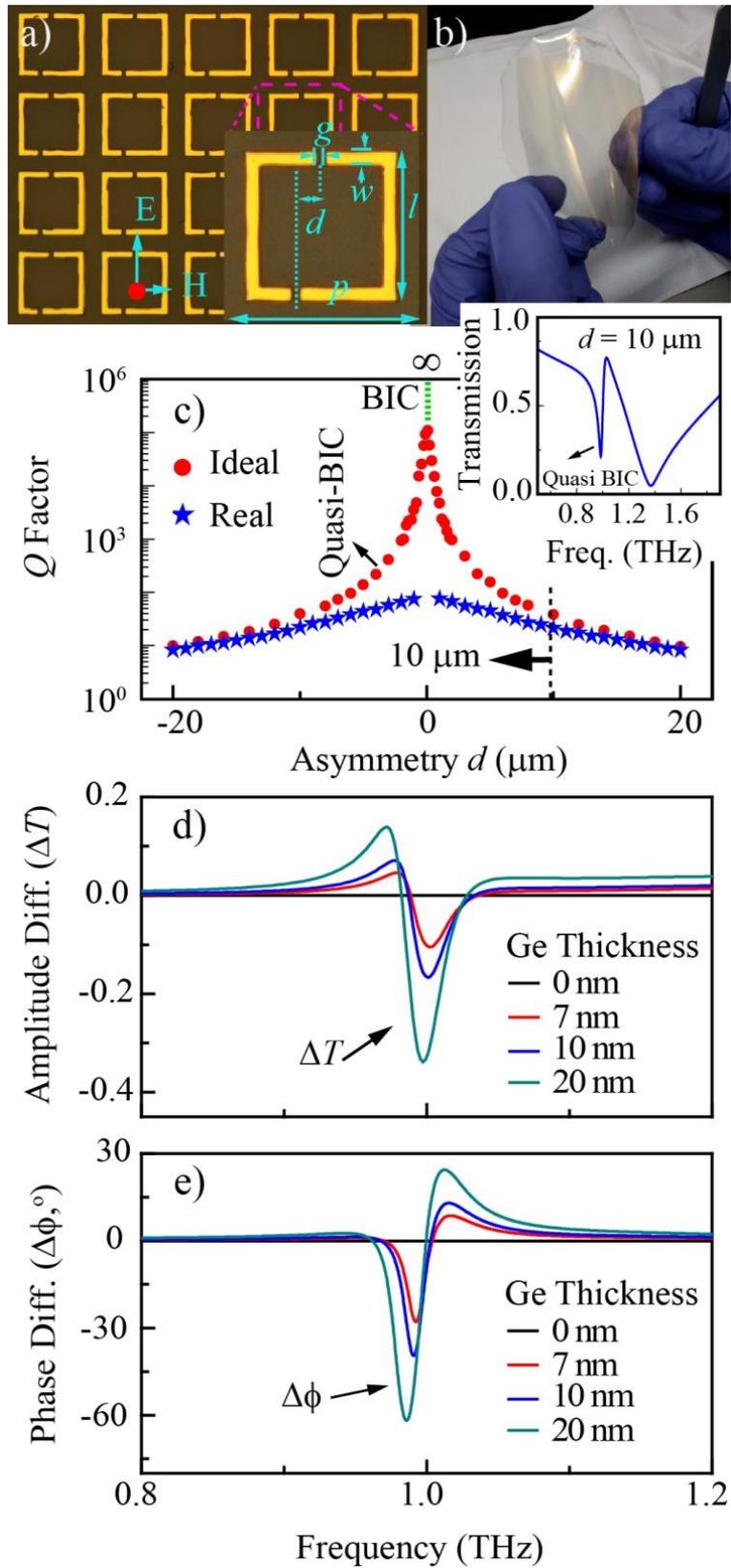

**Figure 1:  Nanometer-scale thin film sensing with quasi-BIC Fano resonance**. a) Optical image of the fabricated TASR metamaterial. Inset depicts the geometrical parameters of the unit cell, gap $g$ = 3 µm, width $w$ = 6 µm, length $l$ = 60 µm, periodicity $p$ = 75 µm and asymmetry $d$ = 10 µm. b) Image of fabricated metamaterial showing robustness and flexibility. c) The $Q$ factors of quasi-BICs of an ideal (dots, PEC) and a realistic (star, metallic) metamaterial array with varying asymmetry $d$. Inset: Simulated transmission spectra of metallic TASR metamaterial with an asymmetry of $d$ = 10 µm. d) and e) Change in the simulated transmission amplitude ($\Delta T$) and phase ($\Delta\phi$, degree) on coating Ge of thicknesses ranging from 7 to 20 nm on TASR metamaterial.

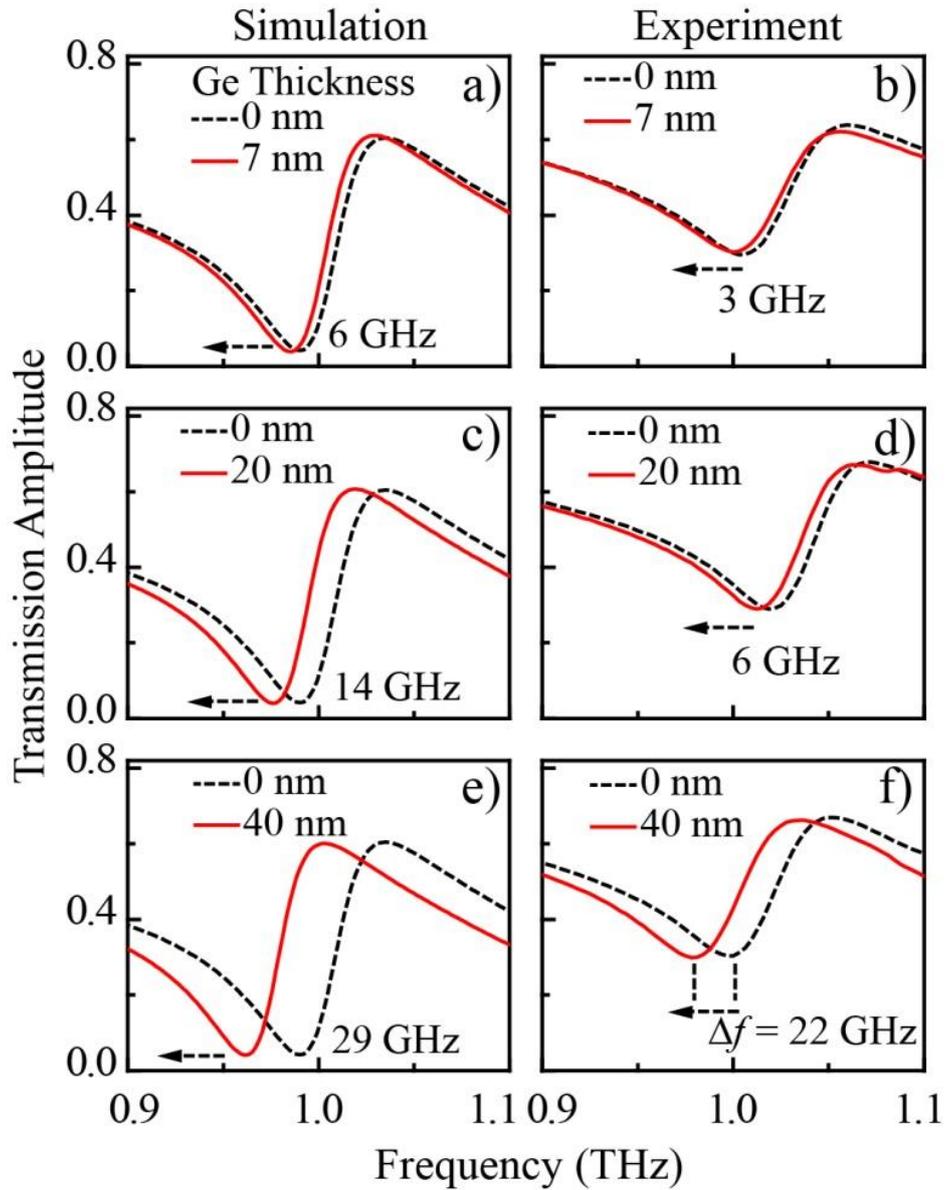

**Figure 2: Simulated and measured transmission amplitude of the flexible metamaterial without and with Ge overlayer.** a), c) and e) depicts the simulated transmission spectra for 7, 20 and 40 nm Ge coating on the metamaterial, respectively. b), d) and f) represents the measured transmission spectra for 7, 20 and 40 nm Ge overlayer on the metamaterial, respectively. The dotted black colour curve represents the transmission spectra of the TASR metamaterial without analyte layer in each case

**Figure 3**

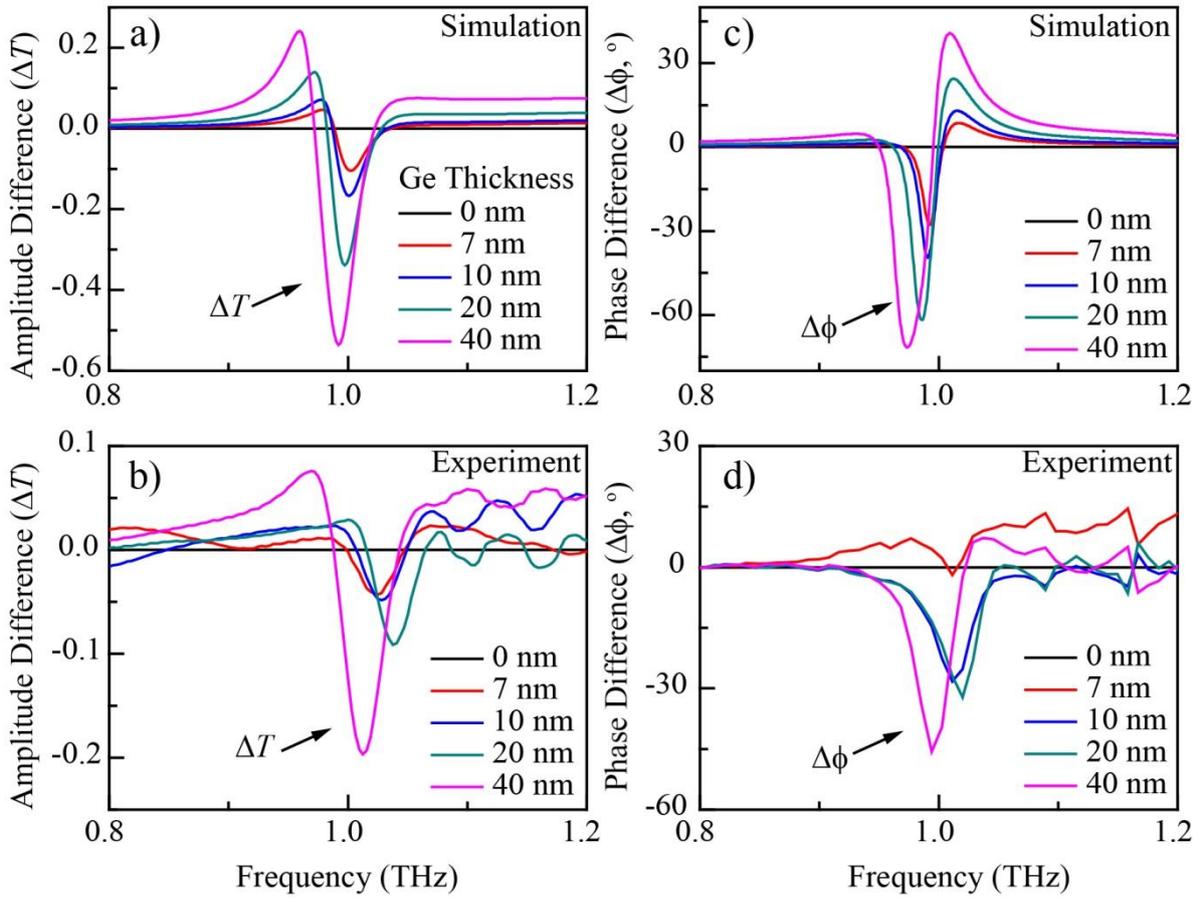

**Figure 3: Change in transmission amplitude (Δ*T*) and phase (Δϕ) spectra for various thicknesses of the Ge overlayer**. a) and b) respectively, represents simulated and experimentally measured transmission amplitude difference for the analyte overlayer of thicknesses ranging from 0 to 40 nm. c) and d), respectively depict the simulated and experimentally measured phase difference for 0 to 40 nm thick analyte.

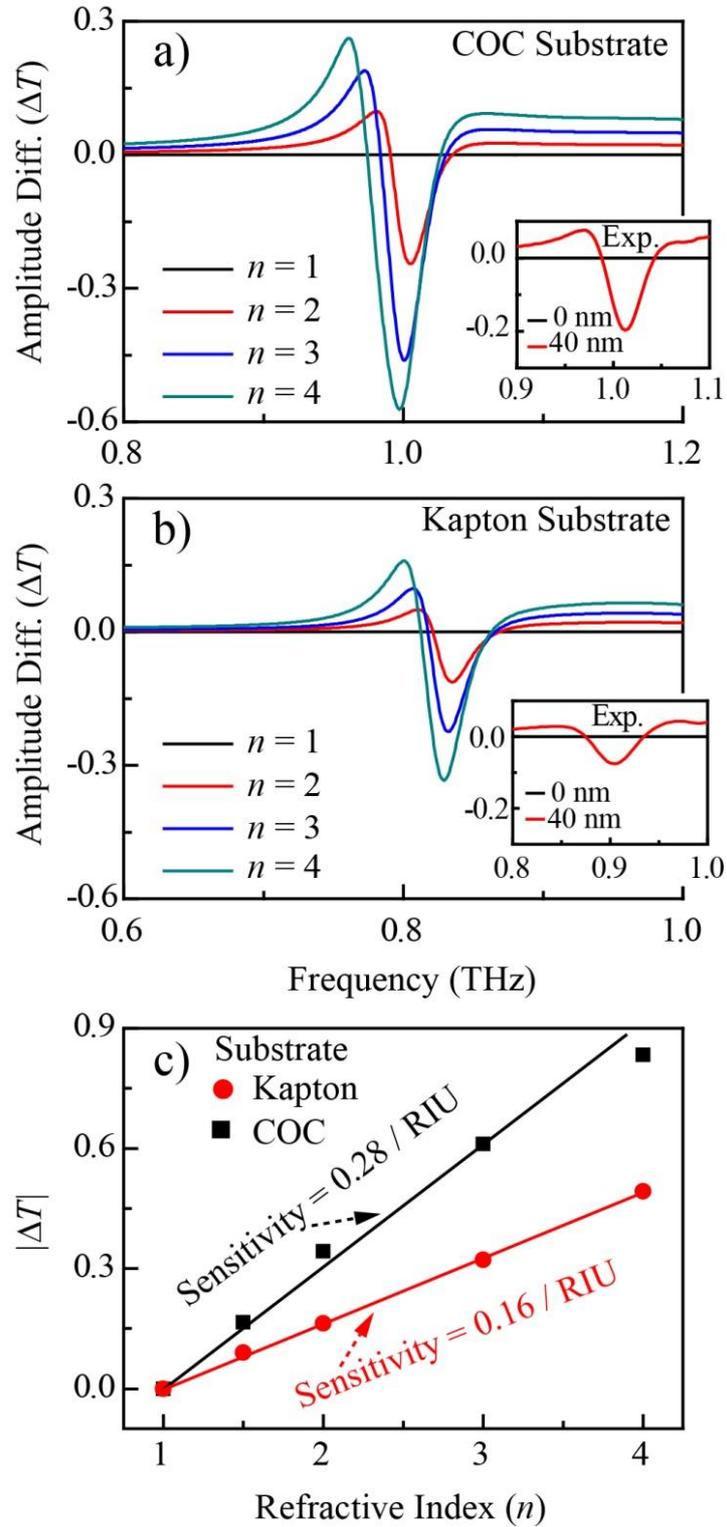

**Figure 4: Sensitivity performance of the metamaterial on different flexible substrates.**

Simulated transmission amplitude difference (Δ*T*) with an analyte of thickness 40 nm but different refractive index placed on the TASR metamaterial residing on a) COC and b) Kapton substrate, with respect to the corresponding uncoated metamaterials. Inset in a) and b) shows experimentally measured $\Delta T$ for 40 nm thick Ge ($n = 4$) on the TASR metamaterials residing on COC and Kapton substrate, respectively. c) Peak-to-peak transmission amplitude difference ($|\Delta T|$) with change in refractive indices of the 40 nm thick analyte placed on TASR metamaterial residing on COC (black) and Kapton (red) substrates.